\documentclass[onecollarge,natbib]{svjour2}
\bibpunct{[}{]}{;}{n}{}{,} % to get "[numbered]" references from natbib
\smartqed  % flush right qed marks, e.g. at end of proof
\usepackage{graphicx}
\usepackage{dcolumn}% Align table columns on decimal point
\usepackage{bm}% bold math
\usepackage{amssymb}
\usepackage{amsfonts}
\usepackage{amsmath}
\usepackage{latexsym}
\usepackage{dsfont}
\usepackage{multirow}
\usepackage{color}
\usepackage[mathscr]{eucal}
\usepackage{subfigure}
\usepackage{bm}
\usepackage{slashed}
\usepackage{multicol}
\usepackage{blindtext}

\newcommand{\ud}{\mathrm{d}}

\newcommand{\uvec}[1]{\vec{#1}}
\newcommand{\pure}{\text{pure}}
\newcommand{\phys}{\text{phys}}

\newcommand{\bea}{\begin{eqnarray}}
\newcommand{\eea}{\end{eqnarray}}
\newcommand{\be}{\begin{equation}}
\newcommand{\ee}{\end{equation}}

\newcommand{\ba}{\begin{eqnarray}}
\newcommand{\ea}{\end{eqnarray}}

\definecolor{green}{rgb}{0,.5,0}

\usepackage[normalem]{ulem}

\renewcommand\sout{\bgroup \color[rgb]{0.55,0.00,0.99} \ULdepth=-.5ex \ULset}

\journalname{Few-Body Systems}
\begin{document}

\title{Quark and gluon orbital angular momentum:\\
Where are we?%\thanks{Grants or other notes
}


\author{C\'edric Lorc\'e         \and
       Keh-Fei Liu %etc.
}


\institute{C. Lorc\'e \at
             Centre de Physique Th\'eorique, \'Ecole polytechnique, CNRS, Universit\'e Paris-Saclay, F-91128 Palaiseau, France \\
              \email{cedric.lorce@polytechnique.fr}           %  \\
           \and
           K.-F. Liu \at
           Department of Physics and Astronomy, University of Kentucky, Lexington, KY 40506, USA
}

\date{Received: date / Accepted: date}

\maketitle

\begin{abstract}
The orbital angular momentum of quarks and gluons contributes significantly to the proton spin budget and attracted a lot of attention in the recent years, both theoretically and experimentally. We summarize the various definitions of parton orbital angular momentum together with their relations with parton distributions functions. In particular, we highlight current theoretical puzzles and give some prospects.
\end{abstract}

\section{Introduction}
\label{sec1}

One of the major challenges in hadron physics is to unravel how the spin of the proton arises from the spin and orbital motion of its constituents. It appears that the quark and gluon spin contributions account for about $60\%$ of the proton spin budget~\cite{Aidala:2012mv,deFlorian:2014yva,Nocera:2014gqa}, implying that $40\%$ should be accounted by the quark and gluon orbital angular momentum (OAM). This is a large fraction which reflects the relativistic nature of the quark-gluon bound state. The quark and gluon OAM, being a correlation between position and momentum, is more difficult to access experimentally than the spin. It depends also on how the total angular momentum (AM) is divided into separate quark and gluon contributions, which is intrinsically ambiguous due to quark-gluon couplings. 

It has long been thought that only the kinetic (or mechanical) decomposition of the proton spin makes sense because its quark and gluon contributions can be extracted from experimental data and computed on the lattice. Recent theoretical and experimental progress have, however, shown that the canonical decomposition can also be accessed experimentally and computed on the lattice, though in a more complicated way. Kinetic and canonical decompositions appear to be complementary with their own advantages and disadvantages. For more detailed discussions, see the recent reviews~\cite{Leader:2013jra,Wakamatsu:2014zza}.

We present here a short summary of the theoretical status of quark and gluon OAM. First we discuss the most general gauge-invariant decomposition of the proton spin and show how it is related to the other ones proposed in the literature. We also comment on its physical significance and stress the importance of the role played by the experimental configuration and the theoretical framework in deciding which explicit form to use. Then we summarize the relations between measurable parton distributions and the different contributions to the proton spin. Having established relations at the integrated level, we finally discuss the status of those proposed at the density level. More details and discussions including recent important lattice developments can be found in Ref.~\cite{Liu:2015xha}

\section{Gauge-invariant decomposition of angular momentum}\label{sec2}

The total AM operator in QCD can be decomposed as
\begin{equation}\label{decomposition}
\uvec J=\uvec S^q+\uvec S^G+\uvec L^q_\text{kin}+\uvec L^G_\text{can}+\uvec L_\text{pot},
\end{equation}
where the quark spin, gluon spin, quark kinetic OAM, gluon canonical OAM and potential OAM contributions are respectively given by
\begin{equation}\label{dec5}
\begin{aligned}
\uvec S^q&=\int\ud^3r\,\psi^\dag\tfrac{1}{2}\uvec\Sigma\psi,&
\uvec S^G&=\int\ud^3r\,\uvec E^a\times\uvec A^a_\phys,\\
\uvec L^q_\text{kin}&=\int\ud^3r\,\psi^\dag(\uvec r\times i\uvec D)\psi,&
\uvec L^G_\text{can}&=-\int\ud^3r\,E^{aj}(\uvec r\times\uvec{\mathcal D}^{ab}_\pure) A^{bj}_\phys,\\
\uvec L_\text{pot}&=-\int\ud^3r\,\rho^a\,\uvec r\times\uvec A^a_\phys
\end{aligned}
\end{equation}
We followed Chen \emph{et al.}~\cite{Chen:2008ag,Chen:2009mr} and decomposed the gauge potential into a pure-gauge field and a ``physical'' field~\cite{Lorce:2013gxa,Lorce:2013bja}
\begin{equation}\label{Adec}
A^\mu=A^\mu_\pure+A^\mu_\phys
\end{equation}
such that $F^{\mu\nu}_\pure=\partial^\mu A^\nu_\pure-\partial^\nu A^\mu_\pure-ig[A^\mu_\pure,A^\nu_\pure]=0$. Under a gauge transformation, $A^\mu_\pure$ and $A^\mu_\phys$ transform as
\begin{equation}
A^\mu_\pure\mapsto U[A^\mu_\pure+\tfrac{i}{g}\partial^\mu]U^{-1},\qquad A^\mu_\phys\mapsto U A^\mu_\phys U^{-1}. 
\end{equation}
The pure-gauge covariant derivatives are defined like the ordinary covariant derivatives with the gauge potential $A^\mu$ replaced by the pure-gauge field $A^\mu_\pure$. This ensures explicit gauge invariance of each contribution in Eq.~\eqref{dec5}.

Due to the QCD equations of motion $\rho^a=g\psi^\dag t^a\psi=\uvec{\mathcal D}^{ab}\cdot\uvec E^b$, the potential OAM can be interpreted as either a quark or a gluon contribution. The quark canonical OAM appearing in the (gauge-invariant) canonical decomposition is obtained by combining quark kinetic and potential OAM
\begin{equation}
\uvec L^q_\text{can}=\uvec L^q_\text{kin}+\uvec L_\text{pot}=\int\ud^3r\,\psi^\dag(\uvec r\times i\uvec D_\pure)\psi.
\end{equation}
The gluon kinetic OAM appearing in the (gauge-invariant) kinetic decomposition is obtained by combining gluon canonical and potential OAM
\begin{equation}
\uvec L^G_\text{kin}=\uvec L^G_\text{can}+\uvec L_\text{pot}=\int\ud^3r\,\uvec r\times[(\uvec A^a_\phys\times\uvec{\mathfrak D}^{ab}_\pure)\times \uvec E^b],
\end{equation}
where $\mathfrak D^\mu_\pure=\tfrac{1}{2}(\mathcal D^\mu+\mathcal D^\mu_\pure)$~\cite{Lorce:2013fpa}.

The decomposition of the gauge potential~\eqref{Adec} essential for ensuring gauge invariance is not unique since 
\begin{equation}
A^\mu_\pure\mapsto A^\mu_\pure+\partial^\mu C\qquad A^\mu_\phys\mapsto A^\mu_\phys-\partial ^\mu C,
\end{equation}
referred to as a Stueckelberg transformation~\cite{Stoilov:2010pv,Lorce:2012rr}, leaves the fundamental Lagrangian invariant. In practice, this is actually not an issue since the experimental conditions combined with the theoretical framework usually provide a natural decomposition. In experiments probing the internal structure of the proton, the off-shell probe indeed provides a natural direction~\cite{Collins:2011zzd} along which one can unambuously define spin and OAM contributions~\cite{Lorce:2012rr,Bashinsky:1998if,Wakamatsu:2014toa}.

\section{Accessing angular momentum with parton distributions}\label{sec3}

Ji~\cite{Ji:1996ek} derived a remarkable relation between the total kinetic AM of quark and gluons, and twist-2 generalized parton distributions (GPDs)
\begin{equation}\label{Jirel}
\langle J^{q,G}_\text{kin}\rangle=\tfrac{1}{2}\int\ud x\,x[H^{q,G}(x,0,0)+E^{q,G}(x,0,0)].
\end{equation}
The kinetic OAM of quarks and gluons can then be obtained by subtracting the corresponding spin contributions, 
\begin{equation}\label{JiOAMrel}
\langle L^{q,G}_\text{kin}\rangle=\langle J^{q,G}_\text{kin}\rangle-\langle S^{q,G}\rangle.
\end{equation}
which are given in the $\overline{MS}$ scheme by the first Mellin moment of the quark and gluon helicity distributions $\langle S^q\rangle=\tfrac{1}{2}\int\ud x\,\Delta q(x)$, $\langle S^G\rangle=\int\ud x\,\Delta g(x)$. Alternatively, it has been shown by Penttinen, Polyakov, Shuvaev and Strikman (PPSS)~\cite{Penttinen:2000dg} that the quark kinetic OAM can also be directly expressed in terms of a two-parton twist-3 GPD~\cite{Kiptily:2002nx,Hagler:2003jw,Hatta:2012cs,Lorce:2015lna}
\begin{equation}\label{twist3rel}
\langle L^q_\text{kin}\rangle=-\int\ud x\,x\,G^q_2(x,0,0).
\end{equation}

Similarly, it is thought that transverse-momentum distributions (TMDs) could also give quantitative information about OAM. Quark model calculations motivated the following relations~\cite{Ma:1998ar,She:2009jq,Avakian:2010br,Efremov:2010cy}
\begin{equation}\label{LzTMD}
\begin{aligned}
\langle L_\text{can}^q\rangle
&=\int\ud x\,\ud^2k_\perp\left[h^q_1(x,\uvec k^2_\perp)-g^q_{1L}(x,\uvec k^2_\perp)\right],\\
&=-\int\ud x\,\ud^2k_\perp\,\tfrac{\uvec k_\perp^2}{2M^2}\,h_{1T}^{\perp q}(x,\uvec k^2_\perp)
\end{aligned}
\end{equation}
but they turned out to be valid only under some restricted condition, and not in general in QCD~\cite{Lorce:2011dv,Lorce:2011zta,Lorce:2011kn}. Alternatively, Burkardt~\cite{Burkardt:2002ks,Burkardt:2003uw} suggested that a chromodynamic lensing mechanism could relate the quark Sivers TMD $f_{1T}^{\perp q}(x,\uvec k^2_\perp)$ and the quark GPD $E^q(x,\xi,t)$. Although this mechnaism can hardly be put on a firm theoretical ground, a variation of it by Bacchetta and Radici~\cite{Bacchetta:2011gx} has led to a new estimate of $\langle J^{q}_\text{kin}\rangle$, in good agreement with most common GPD extractions.

The most natural expression for the OAM is that of a phase-space integral~\cite{Lorce:2011kd,Lorce:2011ni} 
\begin{equation}\label{OAMWigner}
\langle L^{q,G}_{\mathcal W}\rangle=\int\ud x\,\ud^2k_\perp\,\ud^2b_\perp\,(\vec b_\perp\times\vec k_\perp)_z\,\rho^{q,G}_{++}(x,\vec k_\perp,\vec b_\perp;\mathcal W),
\end{equation}
where the Wilson line $\mathcal W$ ensures gauge invariance. In a semi-classical interpretation, the Wigner distribution $\rho^{q,G}_{++}$ gives the (quasi-)probability for finding, at the transverse position $\uvec b_\perp$ a quark or a gluon with momentum $(xP^+,\uvec k_\perp)$ inside a longitudinally polarized ($\Lambda'=\Lambda=+$) proton. This Wigner distribution is related \emph{via} Fourier transform to the generalized transverse-momentum dependent distributions (GTMDs)~\cite{Lorce:2011dv,Meissner:2009ww,Lorce:2013pza}, leading to the simple relation~\cite{Lorce:2011kd,Hatta:2011ku,Kanazawa:2014nha}
\begin{equation}\label{LzGTMD}
\langle L^{q,G}_{\mathcal W}\rangle=-\int\ud x\,\ud^2k_\perp\,\tfrac{\uvec k^2_\perp}{M^2}\,F^{q,G}_{14}(x,0,\uvec k_\perp,\uvec 0_\perp;\mathcal W).
\end{equation}
For a staple-like Wilson line $\mathcal W_{\sqsupset}$~\cite{Collins:2011zzd}, Eq.~\eqref{LzGTMD} gives the canonical OAM $\langle L^{q,G}_\text{can}\rangle=\langle L^{q,G}_{\mathcal W_{\sqsupset}}\rangle$ irrespective of whether the staple is future- or past-pointing~\cite{Hatta:2012cs,Lorce:2015lna,Hatta:2011ku,Ji:2012ba,Lorce:2012ce}. For a straight Wilson line $\mathcal W_\mid$, it gives the kinetic version of the quark OAM $\langle L^q_\text{kin}\rangle=\langle L^q_{\mathcal W_\mid}\rangle$~\cite{Ji:2012sj} and the Ji-Xiong-Yuan (JXY)~\cite{Ji:2012ba} definition of the gauge-invariant gluon OAM $\langle L^G_\text{JXY}\rangle=\langle L^G_{\mathcal W_\mid}\rangle$, where $\uvec L^G_\text{JXY}=-\int\ud^3r\,E^{aj}(\uvec r\times\uvec{\mathcal D}^{ab}) A^{bj}_\phys$. Unfortunately, it is not known so far how to extract them from actual experiments, except possibly at small $x$~\cite{Meissner:2009ww}. Moreover, the GTMD $F_{14}$ does not reduce to any GPD or TMD, and so cannot be directly constrained. However, in the last few years recent developments~\cite{Ji:2014lra,Sufian:2014jma,Zhao:2015kca} opened the interesting possibility of computing GTMDs and OAM directly on the lattice. Moreover, phenomenological models, constrained by experimental data, also provide indirect access to GTMDs and hence OAM.

\section{Angular momentum at the density level}\label{sec4}

Angular momentum can also be defined at the density level and, therefore, be mapped in both position and momentum spaces. Many definitions, differing by superpotential terms of the form $\partial_\alpha X^{[\alpha\mu]\cdots}$, where square brackets stand for antisymmetrization of indices, have been proposed in the literature, creating a somewhat confusing situation. It is essential to keep track of these superpotential terms, since they affect the interpretation of the density.

Hoodbhoy, Ji and Lu~\cite{Hoodbhoy:1998yb} defined higher moments of the kinetic AM densities and concluded that
\begin{equation}\label{kindens}
\langle L^q_\text{kin}\rangle(x)=\tfrac{x}{2}\left[H^q(x,0,0)+E^q(x,0,0)\right]-\tfrac{1}{2}\Delta q(x).
\end{equation}
However, this definition is somewhat \emph{ad hoc} since a complicated tower $\Delta L^{+\cdots +}$ has been added to the natural simple one
\begin{equation}
\tilde L^{q,+\cdots +}_\text{kin}=\tfrac{1}{n+1}\sum_{j=0}^{n}\int\ud^3r\,\overline\psi\gamma^+(iD^+)^j(\uvec r_\perp\times i\uvec D_\perp)_z\,(iD^+)^{n-j}\psi
\end{equation}
in order to have it evolve as a leading-twist operator. Ji, Xiong and Yuan argued that the integrand of the Ji relation~\eqref{Jirel} $\tfrac{x}{2}\left[H^q(x,0,0)+E^q(x,0,0)\right]$ could naturally be identified with an angular momentum decomposition of the transverse component of the quark angular momentum in a transversely polarized target~\cite{Ji:2012ba,Ji:2012sj,Ji:2013tva}. However, a careful inspection revealed several caveats leading to the conclusion that this interpretation is not justified~\cite{Leader:2012md,Harindranath:2012wn,Harindranath:2013goa}.

It is actually not so surprising that the integrand of the Ji relation~\eqref{Jirel} cannot be simply interpreted as the density of parton AM. Indeed, the covariant derivative in the definition of kinetic OAM necessarily implies the contribution of higher-twist parton distributions. H\"agler, Mukherjee and Sch\"afer~\cite{Hagler:2003jw} proposed another density of OAM which now involves both twist-2 and twist-3 GPDs
\begin{equation}\label{HMS}
\langle L^q_\text{kin}\rangle(x)=x\left[H^q(x,0,0)+E^q(x,0,0)+G^q_2(x,0,0)-2G^q_4(x,0,0)\right]-\Delta q(x)
\end{equation}
but it has been obtained within the Wandzura-Wilczek approximation, where the distinction between canonical and kinetic OAM disappears. In a detailed discussion of twist-3 GPDs, Hatta and Yoshida~\cite{Hatta:2012cs} stressed that the density of kinetic OAM is ambiguous because it involves two longitudinal momentum fractions $x_1$ and $x_2$, contrary to canonical OAM. This is related to the fact that contrary to ordinary derivatives,  covariant derivatives do not commute and, therefore, do not admit a unique non-local generalization~\cite{Lorce:2012ce}.

Clearly, canonical OAM is more amenable to a description at the density level. Following the work of Jaffe and Manohar~\cite{Jaffe:1989jz}, a natural density of quark and gluon canonical OAM in the light-front gauge has been defined by Harindranath and Kundu~\cite{Harindranath:1998ve} and later improved by Bashinsky and Jaffe~\cite{Bashinsky:1998if} who provided an explicit expression invariant under residual gauge transformations. Recently, it has been shown that the density of OAM can directly be defined in a gauge-invariant way in phase-space~\cite{Lorce:2011kd,Lorce:2011ni,Lorce:2012ce}
\begin{equation}
\langle L^{q,G}_{\mathcal W}\rangle(x,\uvec k_\perp,\uvec b_\perp)=(\uvec b_\perp\times\uvec k_\perp)_z\,\rho^{q,G}_{++}(x,\uvec k_\perp,\uvec b_\perp;\mathcal W).
\end{equation}
Contrary to the integrated version, the density depends on whether the staple points toward the future $\mathcal W_{\sqsupset}$ or the past $\mathcal W_{\sqsubset}$~\cite{Liu:2015xha,Lorce:2015sqe}. Indeed, using the constraints imposed by parity and time-reversal symmetries, the Wigner distribution of unpolarized quarks and gluons inside a longitudinally polarized proton can be decomposed into four contributions~\cite{Lorce:2015sqe}
\begin{equation}
\rho^{q,G}_{++}=\rho^{q,G}_1+(\uvec b_\perp\cdot\uvec k_\perp)\,\rho^{q,G}_2+(\uvec b_\perp\times\uvec k_\perp)_z\,\rho^{q,G}_3+(\uvec b_\perp\times\uvec k_\perp)_z\,(\uvec b_\perp\cdot\uvec k_\perp)\,\rho^{q,G}_4,
\end{equation}
where $\rho^{q,G}_i\equiv\rho^{q,G}_i(x,\uvec k^2_\perp,(\uvec b_\perp\cdot\uvec k_\perp)^2,\uvec b^2_\perp;\mathcal W)$. The functions $\rho_1$ and $\rho_2$ are related via Fourier transform to the real and imaginary parts of the GTMD $F_{11}$, and similarly for $\rho_3$ and $\rho_4$ with $F_{14}$. The coefficient $(\uvec b_\perp\cdot\uvec k_\perp)$ implies that $\rho_2$ and $\rho_4$ are naive $\mathsf T$-odd, \emph{i.e.} they change sign under $\mathcal W_{\sqsupset}\leftrightarrow\mathcal W_{\sqsubset}$. Integrating over $\uvec b_\perp$ or $\uvec k_\perp$, one is left with just $\rho_3$, \emph{i.e.} the real part of $F_{14}$. On the other hand, a straight Wilson line $\mathcal W_\mid$ leads to the density of JXY quark and gluon OAM~\cite{Ji:2012ba,Ji:2012sj} provided that $\uvec k_\perp$ is integrated over~\cite{Lorce:2012ce}.

The Ji relation~\eqref{Jirel} has also been discussed in position space. Since the information about the spatial distribution of quarks and gluons is encoded in the $t$ dependence of GPDs~\cite{Burkardt:2000za,Burkardt:2002hr}, Polyakov~\cite{Polyakov:2002yz,Goeke:2007fp} suggested that the Ji relation generalized to $t\neq 0$ should provide information about the spatial distribution of kinetic OAM
\begin{equation}
\langle J^{q,G}_\text{kin}\rangle(t)=\tfrac{1}{2}\int\ud x\,x[H^{q,G}(x,0,t)+E^{q,G}(x,0,t)].
\end{equation}
Interestingly, Adhikari and Burkardt~\cite{Adhikari:2013ima} observed within the scalar diquark model with Pauli-Villars regularization that the quark kinetic and canonical OAM spatial densities do not coincide $\langle L^q_\text{kin}\rangle(\uvec b_\perp)\neq\langle L^q_\text{can}\rangle(\uvec b_\perp)$, contrary to their integrated counterparts $\langle L^q_\text{kin}\rangle=\langle L^q_\text{can}\rangle$. It should however be noted that these spatial densities have been defined as the Fourier transform of the $t$-dependent distribution. This is however not justified in all cases. For example, the total AM, is given by the Fourier transform of the combination $\langle J^q_\text{kin}\rangle(t)+t\,\tfrac{\partial}{\partial t}\langle J^q_\text{kin}\rangle(t)$~\cite{Leader:2013jra}, not just $\langle J^q_\text{kin}\rangle(t)$. Note also that the spatial density of quark kinetic OAM has been defined as $\langle L^q_\text{kin}\rangle(\uvec b_\perp)=\langle J^q_\text{kin}\rangle (\uvec b_\perp)-\langle S^q\rangle (\uvec b_\perp)$, but a superpotential ruins this relation at the density level~\cite{Leader:2013jra}. Further investigations are therefore needed.

\section{Summary}\label{sec8}

We briefly reviewed the recent developments about kinetic and canonical decompositions of the proton spin. We discussed in particular the issue of gauge invariance and the link with measurable parton distributions. We also critically commented the various definitions of quark and gluon angular momentum densities.

\end{document}